\documentclass[preprint,pteplogo]{ptephy_v2}%%%%%% to generate preprint number with ptep logo

\preprintnumber{XXXX-XXXX} %%% %%% Insert preprint number here
\usepackage{hyperref}
\usepackage{subcaption}
\usepackage{siunitx}
\usepackage{isotope}
\usepackage{mathrsfs}
\begin{document}

%\title{Establishing the \( \isotope[40]{Ca}(p,p\alpha) \) reaction at \qty{392}{MeV} under quasi-free scattering conditions}

\title{Establishing the \textrm{\(^{\bold{40}}\)Ca(p,p$\boldsymbol{\alpha}$)} reaction at 392~MeV under quasi-free scattering conditions}

%%%% To generate auto affiliation numbers please use \author{}\affil{} command

\author{Riku Matsumura$^{1,2}$}
\author{Junki Tanaka$^{3,4}$\thanks{Corresponding author: \texttt{junki@rcnp.osaka-u.ac.jp}}}
\author{Kazuki Yoshida$^{3}$}
\author{Deuk Soon Ahn$^{5,6}$}
\author{Didier Beaumel$^{7}$}
\author{Jiawei Bian$^{8}$}
\author{Jiawei Cai$^{3}$}
\author{Yoshiki Chazono$^{9}$}
\author{Fengyi Chen$^{4}$}
\author{Masanori Dozono$^{10}$}
\author{Fumitaka Endo$^{3,2}$}
\author{Serge Franchoo$^{7}$}
\author{Tatsuya Furuno$^{11,4}$}
\author{Fumiya Furukawa$^{3}$}
\author{Roman Gernh\"{a}user$^{12}$}
\author{Kevin Insik Hahn$^{5}$}
\author{Jongwon Hwang$^{5}$}
\author{Koshi Higuchi$^{1,2}$}
\author{Yuto Hijikata$^{2,10}$}
\author{Yuya Honda$^{4}$}
\author{Byungsik Hong$^{13}$}
\author{Eiji Ideguchi$^{3}$}
\author{Gen Ikemizu$^{10}$}
\author{Azusa Inoue$^{3,14}$}
\author{Katsuhide Itsuno$^{10}$}
\author{Ryota Iwasaki$^{4}$}
\author{Ryo Kato$^{15}$}
\author{Takahiro Kawabata$^{4}$}
\author{Shoichiro Kawase $^{16}$}
\author{Keita Kawata$^{3}$}
\author{Mukul Khandelwal$^{3}$}
\author{Mingyu Kim$^{13}$}
\author{Sunji Kim$^{5}$}
\author{Nobuyuki Kobayashi$^{3}$}
\author{Yuki Kubota$^{2,17}$}
\author{CheongSoo Lee$^{18}$}
\author{Yutian Li$^{2}$}
\author{Qite Li$^{8}$}
\author{Yifan Lin$^{4}$}
\author{Yukie Maeda$^{15}$}
\author{Yohei Matsuda$^{3,19}$}
\author{Kenjiro Miki$^{20}$}
\author{Maoto Mitsui$^{15}$}
\author{Taichi Miyagawa$^{3}$}
\author{Nikhil Mozumdar$^{21,22}$}
\author{Motoki Murata$^{3}$}
\author{Tomoya Nakada$^{10}$}
\author{Hide Nakama$^{15}$}
\author{Geonhee Oh$^{18}$}
\author{Kazuyuki Ogata$^{9}$}
\author{Shoya Ogawa$^{9}$}
\author{Shingo Ogio$^{10}$}
\author{Shinsuke Ota$^{3}$}
\author{Stefanos Paschalis$^{23}$}
\author{Marina Petri$^{23}$}
\author{Thomas Pohl$^{2}$}
\author{Futa Saito$^{19}$}
\author{Soki Sakajo$^{4}$}
\author{Yohei Sasagawa$^{4}$}
\author{Takafumi Sato$^{15}$}
\author{Hiroaki Shibakita$^{4}$}
\author{Hideya Sonoda$^{15}$}
\author{Taiki Sugiyama$^{1,2,17}$}
\author{Yumaro Suzuki$^{4}$}
\author{Atsushi Tamii$^{3,4}$}
\author{Ryotaro Tsuji$^{2,17,10}$}
\author{Stefan Typel$^{21,24}$}
\author{Satoshi Umemoto$^{19}$}
\author{Xuan Wang$^{3}$}
\author{Cheng Wang$^{8}$}
\author{Guo Wenhao$^{4}$}
\author{Matthew Whitehead$^{23}$}
\author{Riku Yamamoto$^{15}$}
\author{Nobuhiro Yamasaki$^{19}$}
\author{Shunpei Yamazaki$^{25}$}
\author{Zaihong Yang$^{8}$}
\author{Takayuki Yano$^{2,10}$}
\author{Kohki Yasumura$^{19}$}
\author{Ryosuke Yoshida$^{2,10}$}
\author{Jichao Zhang$^{3}$}
\author{Kaijie Zhou$^{8}$}
\author{Juzo Zenihiro$^{10}$}
\author{Tomohiro Uesaka$^{2,17,1}$}

\affil{$^1$Graduate School of Science and Engineering, Saitama University, Saitama 338-8570, Japan}
\affil{$^2$RIKEN Nishina Center for Accelerator-Based Science, Saitama 351-0198, Japan}
\affil{$^3$Research Center for Nuclear Physics (RCNP), The University of Osaka, Osaka 567-0047, Japan}
\affil{$^4$Department of Physics, The University of Osaka, Osaka 560-0043, Japan}
\affil{$^5$Center for Exotic Nuclear Studies, Institute for Basic Science, Daejeon 34126, Republic of Korea}
\affil{$^6$Facility for Rare Isotope Beams, Michigan State University, Michigan 48824, USA}
\affil{$^{7}$Ir\'ene Joliot-Curie Laboratoire de Physique des 2 infinis and University Paris-Saclay, 91405 Orsay, France}
\affil{$^8$School of Physics, Peking University, Beijing 100871, China}
\affil{$^{9}$Department of Physics, Kyushu University, Fukuoka 819-0395, Japan}
\affil{$^{10}$Department of Physics, Kyoto University, Kyoto 606-8502, Japan}
\affil{$^{11}$Department of Applied Physics, University of Fukui, Fukui 910-8507, Japan}
\affil{$^{12}$Physik Department, Technische Universit\"at M\"unchen, Garhing 85748, Germany}
\affil{$^{13}$Department of Physics, Korea University, Seoul 02841, Republic of Korea}
\affil{$^{14}$Department of Physics, University of Oslo, 0316 Oslo, Norway}
\affil{$^{15}$Department of Applied Physics, University of Miyazaki, Miyazaki 889-2192, Japan}
\affil{$^{16}$Department of Advanced Energy Science and Engineering, Kyushu University, Fukuoka 816-8580, Japan}
\affil{$^{17}$RIKEN Pioneering Research Institute, Saitama 351-0198, Japan}
\affil{$^{18}$Institute for Rare Isotope Science, Institute for Basic Science, Daejeon 34000, Republic of Korea}
\affil{$^{19}$Department of Physics, Konan University, Kobe 658-8501, Japan}
\affil{$^{20}$Department of Physics, Tohoku University, Miyagi 980-8578, Japan}
\affil{$^{21}$Institut f\"ur Kernphysik, Technische Universit\"at Darmstadt, 64289 Darmstadt, Germany}
\affil{$^{22}$Helmholtz Forschungsakademie Hessen f\"ur FAIR, 64291 Darmstadt, Germany}
\affil{$^{23}$School of Physics, Engineering and Technology, University of York, York, United Kingdom}
\affil{$^{24}$GSI Helmholtzzentrum f\"ur Schwerionenforschung, Planckstra\ss e 1, 64291 Darmstadt, Germany}
\affil{$^{25}$Research Center for Accelerator and Radioisotope Science(RARiS),Sendai 980-8578, Japan}

\begin{abstract}
The \( (p,p\alpha) \) reaction offers a direct means to probe preformed \( \alpha \)-cluster structures in nuclei under quasi-free scattering conditions.
Previous studies around \qty{100}{MeV} provided valuable insights into \( \alpha \) clustering, but quantitative comparison with microscopic cluster wave functions remained limited due to strong distortion effects.
At higher energies, the reaction mechanism becomes simpler and the distorted-wave impulse approximation (DWIA) provides a more reliable framework for quantitative analysis.
In the present work, the \isotope[40]{Ca}\( (p,p\alpha) \) reaction was measured at an incident energy of \qty{392}{MeV} using the high-resolution Grand Raiden and LAS spectrometers at RCNP.
Despite the small cross section in this energy region, the achieved resolution allowed clear separation of the ground and excited states of the residual \isotope[36]{Ar} nucleus, and corresponding momentum distributions were extracted.
%These results provided the first direct evidence of the \( \alpha \)-cluster component in \isotope[40]{Ca} at high energy.
DWIA calculations using a Woods--Saxon \( \alpha + \isotope[36]{Ar} \) bound-state wave function yielded an experimental spectroscopic factor of \( S_{\mathrm{FAC}}^{\mathrm{WS}} = 0.51 \pm 0.05 \), consistent with the previous result at \qty{101.5}{MeV} (\(0.52 \pm 0.23 \)).
This agreement demonstrates that the reaction mechanism is well described across a wide energy range.
The present study establishes the feasibility of high-precision \( (p,p\alpha) \) measurements at several hundred MeV and highlights their potential as a quantitative probe of \( \alpha \) clustering in medium-mass nuclei, forming the basis for systematic studies in both stable and unstable systems.
\end{abstract}

\subjectindex{xxxx, xxx}

\maketitle
\section{Introduction: Quasi-Free \texorpdfstring{$(p,p\alpha)$}{(p,palpha)} Studies of \texorpdfstring{$\alpha$}{alpha} Clustering}

The \( (p,p\alpha) \) reaction has been developed as a versatile tool to investigate \( \alpha \)-cluster structures in nuclei under quasi-free scattering conditions~\cite{Cowley2021,Yoshida2025}.
In this reaction, a proton incident on the target nucleus knocks out a preformed \( \alpha \) cluster, allowing the direct extraction of its spectroscopic strength and momentum distribution.
Since the outgoing proton and \( \alpha \) particle carry information on the internal motion of the cluster, the \( (p,p\alpha) \) process provides a powerful means of probing \( \alpha \) correlations near the nuclear surface, where clustering is expected to be enhanced.

Early systematic studies were conducted with proton beams of about \qty{100}{MeV} on light nuclei such as \isotope[6,7]{Li}, \isotope[9]{Be}, and \isotope[12]{C}~\cite{Gottschalk1970,Roos1977}.
These experiments established the basic framework of the \( (p,p\alpha) \) reaction, demonstrating that coincident detection of the scattered proton and the ejected \( \alpha \) particle enables the separation of distinct final states and a quantitative extraction of spectroscopic information.
The approach was later extended to medium-mass systems such as \isotope[40]{Ca}~\cite{Carey1981,Carey1984}, marking a significant step toward exploring the evolution of \( \alpha \) clustering beyond the \( p \)-shell region.
By detecting the scattered proton and the knocked-out \( \alpha \) particle in coincidence, information on the existence probability (spectroscopic strength) and the internal momentum distribution of \( \alpha \) clusters has been obtained~\cite{Gottschalk1970,Nadasen1989, Nadasen1999,Neveling2008,Bachelier1976,Mabiala2009,Yoshimura1998}.
Although most of these measurements were performed around \qty{100}{MeV}, an early attempt at a much higher energy of \qty{600}{MeV} was also reported~\cite{Landaud1978}, providing a first exploration of the reaction in the several-hundred-MeV region where the impulse approximation is expected to be more valid.

At incident energies near \qty{100}{MeV}, relatively large cross sections allowed measurements with sufficient statistics, making it possible to perform detailed analyses of \( \alpha \)-knockout reactions~\cite{Carey1981,Carey1984}.
In light nuclei such as \isotope[12]{C}, the factorization between the elementary \( p\text{--}\alpha \) scattering amplitude and the nuclear \( \alpha \)-cluster wave function was found to hold approximately under quasi-free scattering (QFS) conditions, allowing the measured cross sections to be consistently interpreted in terms of single-step knockout processes~\cite{Mabiala2009}.
For medium-mass nuclei such as \isotope[40]{Ca}, several experiments have also been performed in this energy region.
A high-resolution measurement at \qty{101.5}{MeV} using the Maryland Cyclotron provided detailed \( \alpha \)-separation energy spectra and energy distributions of the outgoing proton, from which experimental spectroscopic factors were extracted through DWIA analyses~\cite{Carey1981,Carey1984,Nadasen1981,Nadasen1989}.
Subsequent measurements at iThemba LAB examined the analyzing powers and spin observables for \( (p,p\alpha) \) reactions at similar energies, showing that, with carefully chosen optical potentials, the data can be consistently described within the impulse approximation framework~\cite{Neveling2008}.
These results collectively demonstrate that \( (p,p\alpha) \) studies around \qty{100}{MeV} have provided valuable insights into the \( \alpha \)-cluster component of nuclear wave functions, while minor deviations from DWIA predictions at higher ejectile energies suggest the need for a more refined treatment of refraction and absorption effects, particularly in kinematic regions corresponding to large proton energies and small \( \alpha \) scattering angles.

Despite the success at \qty{100}{MeV}, measurements in this energy region remain affected by strong final-state interactions and substantial distortion of the reaction waves, in particular the \( \alpha \)-residue scattering at tens of~MeV in the final state, which obscure direct connections between the experimental cross sections and the microscopic cluster wave functions.
At higher incident energies, on the other hand, these effects become significantly reduced.
High-energy \( (p,p\alpha) \) experiments therefore offer unique advantages.
In the several-hundred-MeV region, the reaction mechanism is greatly simplified, distorting potentials become less absorptive, and the quasi-free condition is more ideally realized.
The higher outgoing energies of the proton and \( \alpha \) particle also expand the accessible kinematic domain and minimize multi-step contributions, leading to a more transparent interpretation of the observed cross sections in terms of direct \( \alpha \) knockout processes.
In addition, final-state interactions (FSI) between the outgoing particles and the residual nucleus are strongly suppressed at higher energies, allowing clearer identification of the direct reaction mechanism and more reliable comparison with theoretical models based on the distorted-wave impulse approximation (DWIA).

Nevertheless, the experimental realization of high-energy \( (p,p\alpha) \) measurements has long remained challenging.
As the incident energy increases, the reaction cross section decreases rapidly, requiring high-intensity proton beams and efficient coincidence detection systems to obtain sufficient statistics.
At the same time, high energy and angular resolutions are essential to resolve closely spaced excited states of the residual nucleus.
For example, a measurement at \qty{157}{MeV} reported the first attempt to extend \( (p,p\alpha) \) spectroscopy to higher energies, but the resolution was insufficient to clearly separate the ground and excited states of the residual \( \isotope[36]{Ar} \) nucleus.
Another experiment at \qty{600}{MeV} achieved access to large momentum transfers, but the low yield limited quantitative analyses~\cite{Landaud1978}.
These results highlight a contrast: while \( \sim \qty{100}{MeV} \) measurements achieved good resolution and statistics, experiments at higher energies have been constrained by both yield and resolution, leaving only a few successful examples above \qty{200}{MeV}~\cite{Yoshimura1998}.

In the present study, to achieve the \( (p,p\alpha) \) reaction at several hundred MeV, a high-intensity \qty{392}{MeV} proton beam was combined with the high-resolution double-arm magnetic spectrometer system at RCNP, and a thin target was employed to optimize the balance between resolution and yield.
As a result, efficient coincidence detection and precise energy determination of the outgoing particles were achieved, providing sufficient resolution to separate individual excited states of the residual nucleus while maintaining high statistical accuracy.
The \isotope[40]{Ca} nucleus serves as an ideal benchmark for such studies, as it exhibits both shell closure and a well-developed \( \alpha \)-cluster component.
By comparing the results at \qty{392}{MeV} with the well-established \qty{101.5}{MeV} data~\cite{Carey1981,Carey1984}, we examine the consistency of the extracted experimental spectroscopic factors across widely different energy regimes, providing a stringent test of the DWIA reaction framework.
Through this comparison, the present work establishes a firm experimental foundation for quantitative \( (p,p\alpha) \) studies at several hundred MeV and demonstrates the feasibility of high-resolution, high-statistics measurements in medium-mass nuclei.
The results pave the way for systematic investigations of \( \alpha \) clustering in both stable and radioactive nuclei under quasi-free conditions, linking the spectroscopic strength observed in knockout reactions to the microscopic structure of preformed \( \alpha \) clusters at the nuclear surface.

In recent years, interest in \( \alpha \) clustering in the ground states of nuclei has been revitalized by the discovery of surface clustering phenomena across a wide mass range, including both stable and neutron-rich systems~\cite{Typel2014,Uesaka2024}.
Systematic \( (p,p\alpha) \) measurements have been initiated on medium-mass and heavy nuclei to explore the mass and isotopic dependence of \( \alpha \) correlations~\cite{Kubota2025}.
For instance, \( (p,p\alpha) \) reactions on tin isotopes have been performed at \qty{400}{MeV} using the double-arm magnetic spectrometer setup, providing the first comprehensive investigation of \( \alpha \)-cluster formation in tin isotopes with increasing neutron number~\cite{Tanaka2021}.
These pioneering studies revealed valuable systematics of clustering phenomena; however, due to the need for thick targets (\qty{40}{mg/cm^2}) to secure sufficient yield, the achievable energy resolution was limited, and uncertainties remained in the absolute normalization of the proton energy spectra, partly reflecting incomplete acceptance calibration.
Consequently, detailed state-by-state analyses could not be performed in those measurements.

\section{Experimental Setup with the Double-Arm Spectrometer System}

\begin{figure}[htbp]
   \centering
   \includegraphics[width=16.5cm]{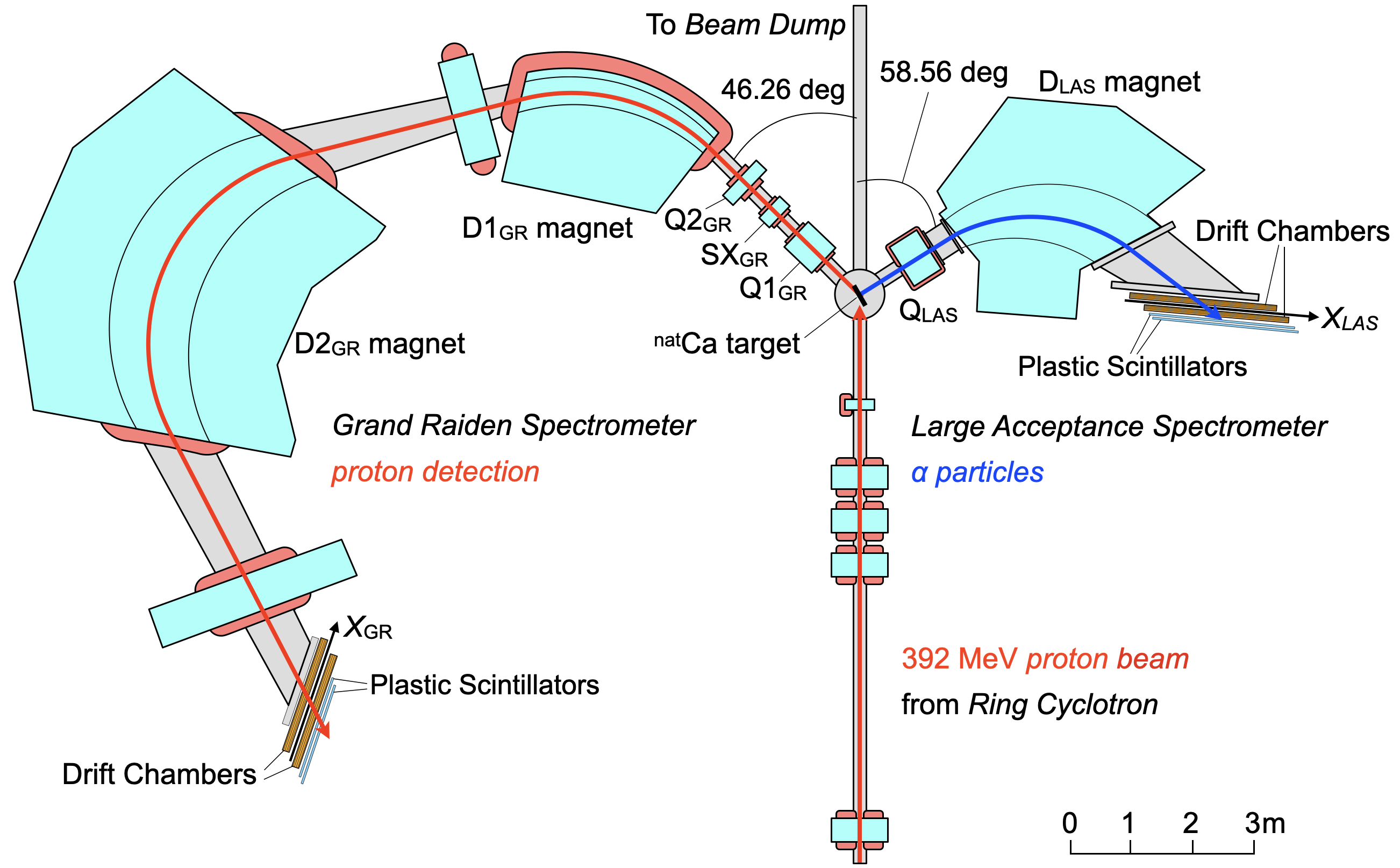}
   \caption{
   Schematic layout of the experimental setup with the double-arm spectrometer used for the \( (p,p\alpha) \) measurement at RCNP.
   }
   \label{fig_setup}
\end{figure}

The experiment was carried out at the Ring Cyclotron Facility of the Research Center for Nuclear Physics (RCNP), The University of Osaka.
A \qty{392}{MeV} proton beam was accelerated by the ring cyclotron and transported through the WS beam line onto a self-supporting \isotope[\text{nat}]{Ca} target.
The scattered proton and the \( \alpha \) particle produced in the \( (p,p\alpha) \) reaction were detected in coincidence and analyzed using the double-arm spectrometer system~\cite{Wakasa2017} consisting of the Grand Raiden (GR)~\cite{Fujiwara1999} and the Large Acceptance Spectrometer (LAS)~\cite{Matsuoka1995}, equipped with their focal-plane detectors.
This setup enabled event-by-event particle identification and precise determination of the momenta under the selected kinematic condition.
A schematic layout of the experimental setup is shown in Fig.~\ref{fig_setup}.

Data were acquired with beam intensities of \qty{50}{nA} for \qty{4}{h} and \qty{100}{nA} for an additional \qty{4}{h}.
The \isotope[\text{nat}]{Ca} target had a thickness of \qty{11.8}{mg/cm^2}.
The foil was mounted perpendicular to the axis pointing from the reaction point toward the LAS spectrometer in order to minimize the energy loss of low-energy \( \alpha \) particles emitted in the LAS direction.
The proton beam was delivered with an energy spread smaller than \( \sim \qty{200}{keV} \) in FWHM and a spot size of approximately \qty{1}{mm} at the target position, ensuring precise definition of the reaction kinematics.
The natural isotopic abundance of the \isotope[40]{Ca} in the \isotope[\text{nat}]{Ca} target is \qty{96.9}{\%}.
The effect of the finite target thickness on the overall energy resolution was evaluated to be about \qty{0.4}{MeV} in sigma, mainly due to angular straggling and energy-loss differences of the outgoing particles.
The unreacted beam was transported to a Faraday cup located at the end wall of the experimental hall, where the beam current was monitored.

The coincidence measurement of the outgoing proton and \( \alpha \) particle was performed using GR and LAS.
The spectrometers were positioned at laboratory polar angles of \qty{46.26}{\degree} (GR) and \qty{58.56}{\degree} (LAS).
These settings correspond to the \( p\text{--}\alpha \) scattering angle of \qty{60}{\degree} in the center-of-mass frame with recoil-less conditions when the reaction \( Q \) value of \qty{-7.038}{MeV} is taken into account.
This angular configuration was chosen to closely match the kinematics of free two-body proton–\( \alpha \) elastic scattering at 392 MeV, taking into account the \( \alpha \) separation energy.
As a result, the momentum transfer is shared between the outgoing proton and the \( \alpha \) particle, minimizing the recoil momentum of the residual nucleus.
The central kinetic energies of the analyzed particles were \qty{318.59}{MeV} for protons and \qty{66.37}{MeV} for \( \alpha \) particles.
The corresponding central magnetic fields were \qty{0.9293}{T} (GR) and \qty{0.6729}{T} (LAS).
After being bent by the spectrometer magnets, the incident particles were dispersed according to their momentum in the horizontal (dispersive) direction.
To enable reliable reconstruction of the scattering angles at the target position, a dedicated sieve-slit calibration was performed.
A sieve slit with a regular array of holes was placed downstream of the target, providing particles with well-defined horizontal and vertical emission angles.
By analyzing the correlations between the known geometrical positions of the sieve holes and the focal-plane observables under the same kinematic and optical conditions as the \( (p,p\alpha) \) measurement, the ion-optical coefficients relevant for scattering-angle reconstruction were determined.
Based on this calibration, the GR Q1 magnet field was reduced by \qty{10}{\%} and the LAS Q magnet field was increased by \qty{20}{\%}, ensuring sufficient sensitivity to the vertical scattering angle in the present experiment.
Details of the sieve-slit calibration and ion-optical analysis are given in Ref.~\cite{Miyagawa2025}.

At the GR focal plane, two sets of multi-wire drift chambers and two plastic scintillators (each \qty{1}{cm} thick) were installed.
The drift chambers provided trajectory tracking, yielding the horizontal (momentum-dispersive) position and angle.
The plastic scintillators covered the entire focal plane horizontally and were read out from both ends by photomultiplier tubes (PMTs); end-to-end coincidences improved the signal-to-noise ratio.
The first scintillator layer provided the trigger, while the second layer yielded energy-loss information.

At the LAS focal plane, two sets of drift chambers and a thin (\qty{3}{mm}) plastic scintillator were installed.
As in GR, the drift chambers provided position and angle information, and the thin scintillator, read out from both ends, generated the trigger in coincidence.
The LAS energy acceptance for \( \alpha \) particles was \qtyrange{50}{90}{MeV}.
The lowest-energy \( \alpha \) particles reached the scintillator with an energy down to \( \sim \qty{5}{MeV} \); to avoid efficiency loss, the PMT gain was increased and the trigger threshold was set as low as practical.
Because the large-area thin scintillator exhibits significant light attenuation, it is not optimal for precise energy-loss measurements; instead, the \( \alpha \) energy loss was effectively inferred from time-over-threshold (TOT) signals obtained from the drift chambers (see, e.g., Ref.~\cite{Bashkirov1999}).
The spectrometer settings and focal-plane detector configurations are summarized in Table~\ref{tab:settings}.
\begin{table}[htbp]
\centering
\caption{Magnetic-field settings and focal-plane detector configurations of GR and LAS. Software acceptances reflect the analysis cuts applied.}
\label{tab:settings}
\begin{tabular}{lcc}
\hline\hline
                                      & GR                            & LAS                          \\
\hline
Detected particle                     & proton                        & \( \alpha \) particle        \\
Central polar angle                   & \qty{46.26}{\degree}           & \qty{58.56}{\degree}          \\
Polar-angle acceptance (software)     & \qty{\pm 1.15}{\degree}        & \qty{\pm 3.15}{\degree}       \\
Central azimuthal angle               & \qty{0}{\degree}               & \qty{180}{\degree}            \\
Azimuthal-angle acceptance (software) & \qty{\pm 2.30}{\degree}        & \qty{\pm 3.32}{\degree}       \\
Solid angle (software)                & \qty{2.5}{msr}                 & \qty{10.0}{msr}               \\
Central kinetic energy                & \qty{318.59}{MeV}              & \qty{66.37}{MeV}              \\
Energy acceptance (software)          & \qtyrange{303}{329}{MeV}       & \qtyrange{50}{80}{MeV}        \\
Magnetic field                        & \qty{0.9293}{T}                & \qty{0.6729}{T}               \\
Field adjustment                      & Q1: \qty{-10}{\%}              & Q: \qty{+20}{\%}              \\
Exit window                           & polyimide \qty{125}{\micro m}  & polyimide \qty{125}{\micro m} \\
Drift chambers                        & 2 sets                        & 2 sets                       \\
Plastic scintillators                 & 2 layers (\qty{1}{cm} each)    & 1 layer (\qty{3}{mm})         \\
Trigger                               & 1st scintillator              & thin scintillator            \\
Energy-loss observable                & 2nd scintillator light output & drift-chamber TOT            \\
\hline\hline
\end{tabular}
\end{table}
\section{Data Analysis}
\subsection{Particle identification and event selection}
\begin{figure}[htbp]
   \centering
   \begin{subfigure}[b]{0.48\linewidth}
      \centering
      \includegraphics[width=\linewidth]{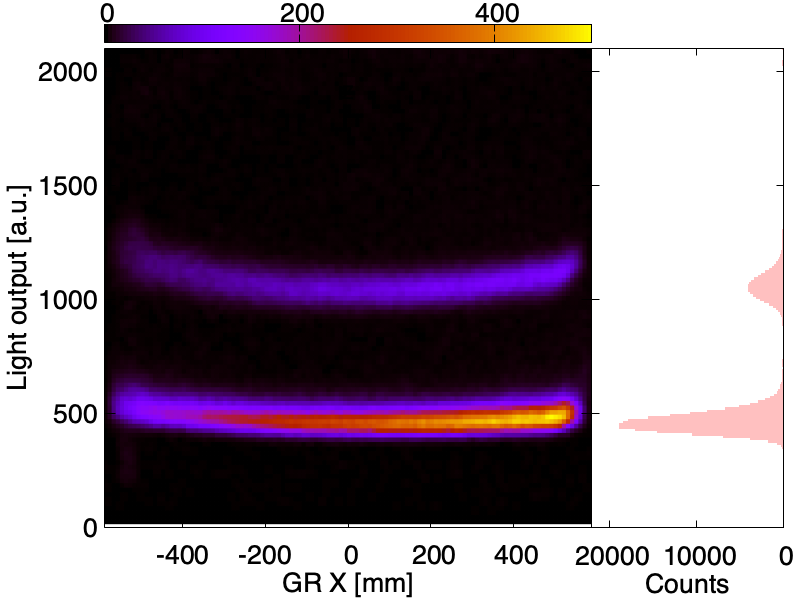}
      \caption{GR}
      \label{fig2a}
   \end{subfigure}
   \hfill
   \begin{subfigure}[b]{0.48\linewidth}
      \centering
      \includegraphics[width=\linewidth]{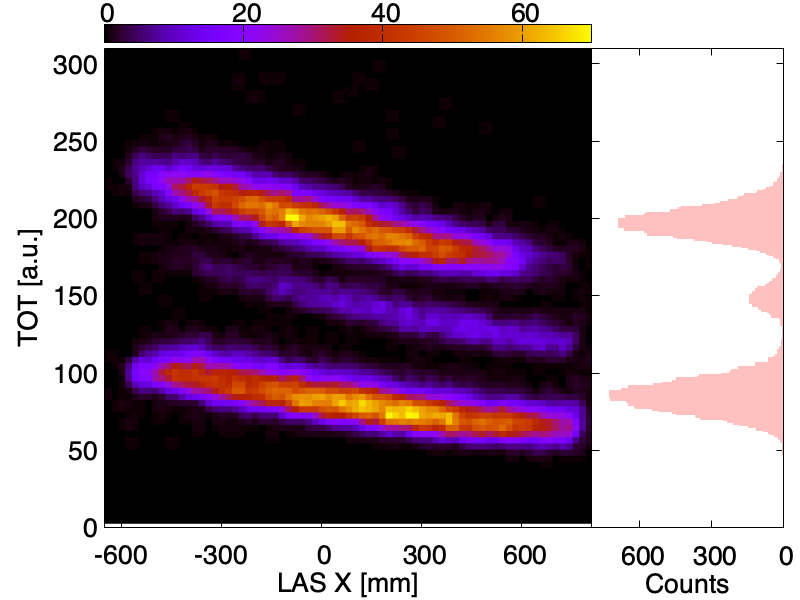}
      \caption{LAS}
      \label{fig2b}
   \end{subfigure}
   \caption{
   Particle identification at the focal planes of GR (a) and LAS (b).
   In each panel, the left plot shows the two-dimensional correlation used for particle identification (horizontal position vs. light output in GR, or horizontal position vs. TOT in LAS), while the right plot displays the corresponding projected spectrum after correcting for the position dependence.
   Distinct bands of protons, deuterons, and \( \alpha \) particles are clearly identified.
   }
   \label{fig2}
\end{figure}

The scattered proton and the knocked-out \( \alpha \) particle from the \isotope[40]{Ca}\( (p,p\alpha) \) reaction were identified at their respective focal-plane detectors.
In GR, particle identification was performed using the correlation between the light output in the second plastic scintillator and the horizontal position (GR X, momentum-dispersive direction) obtained from drift-chamber tracking, as shown in the left panel of Fig.~\ref{fig2}(a).
The light output was defined as the geometric mean of the light-output signals from the left and right photomultiplier tubes (PMTs) of the plastic scintillator, in order to remove the position dependence along the GR focal plane (\( X \)) direction.
The distribution appearing around a light output of 500 corresponds to protons, while a separate distribution above it originates from deuterons.

In the right panel of Fig.~\ref{fig2}(a), the light output is presented as a one-dimensional spectrum.
For this purpose, the residual dependence of the light output on the GR focal-plane position, which was not fully removed by the geometric mean, is corrected.
Specifically, for the proton locus shown in the left panel, a quadratic function was fitted to reproduce this position dependence, and an event-by-event correction was applied to remove it.
Protons were selected within \( \pm 3\sigma \) in this spectrum, and the contamination from deuterons was estimated to be less than \qty{0.1}{\%}.

In LAS, particle identification was carried out using the correlation between the focal-plane position \( X \) and the time-over-threshold (TOT) signals from the drift chambers, as shown in the left panel of Fig.~\ref{fig2}(b).
Distributions corresponding to protons, deuterons, and \( \alpha \) particles were observed.
The right panel of Figure~\ref{fig2}(b) shows the TOT spectrum after correcting for the dependence on LAS X.
The \( \alpha \) particles were selected within \( \pm 2\sigma \) in this spectrum, and the contamination from deuterons was estimated to be \qty{0.5}{\%} of the total counts.

In addition, the detection efficiencies of the plastic scintillators and drift chambers were evaluated.
The detection efficiency associated with multi-hit handling in the plastic scintillators was \qty{99}{\%} for GR and \qty{94}{\%} for LAS, while the tracking efficiency of the drift chambers was \qty{93}{\%} for protons in GR and \qty{98}{\%} for \( \alpha \) particles in LAS.
The particle-identification efficiency together with these detection efficiencies were included in the cross-section normalization and contribute to the overall systematic uncertainty at the level of \qty{4}{\%}.
\subsection{Coincidence timing and accidental background estimation}
\begin{figure}[htbp]
\centering
   \centering
   \includegraphics[width=0.75\linewidth]{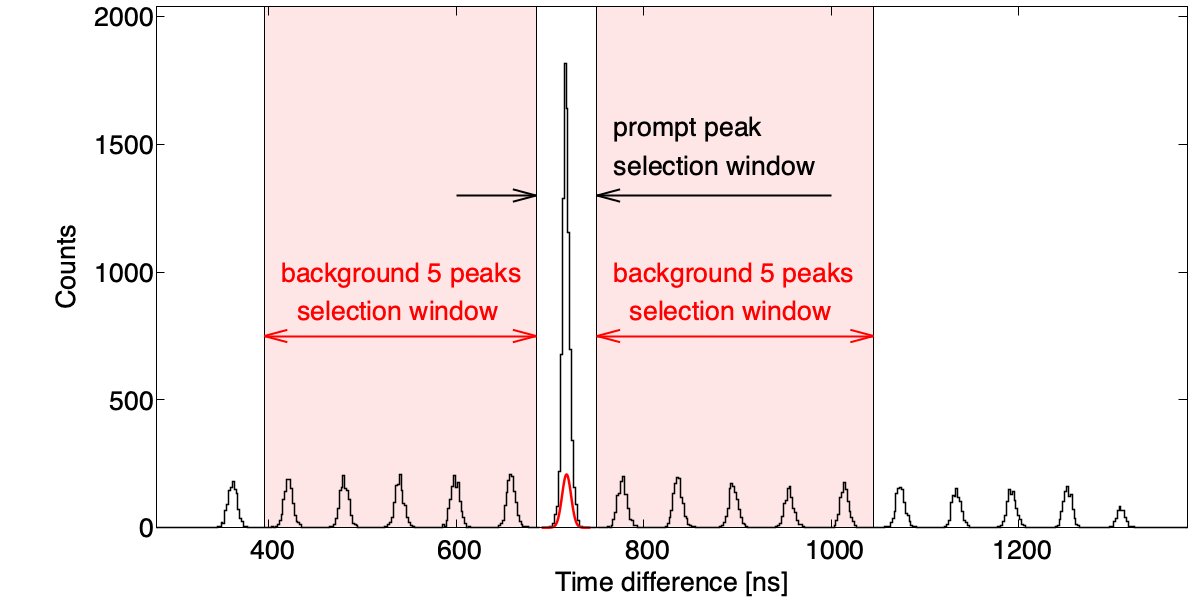}
   \caption{
   Time-difference spectrum between the timing signals of GR and LAS.
   Multiple peaks appear, reflecting the bunch structure of the cyclotron.
   The prompt-peak selection window and background-selection windows are indicated.
   The red line represents the contribution of accidental coincidences evaluated from the side peaks.}
   \label{fig3}
\end{figure}

Figure~\ref{fig3} shows the time-difference spectrum between the timing signals obtained from the trigger plastic scintillators at the focal planes of GR and LAS.
Multiple peaks reflect the bunch structure associated with the acceleration cycle of the cyclotron.
In this spectrum, the protons and \( \alpha \) particles have already been identified in GR and LAS, respectively.
The prompt peak contains both true and accidental coincidence events.
In contrast, the other peaks consist only of accidental coincidences, indicating that accidental coincidences are constantly observed at a uniform rate.
Here, to evaluate the contribution of accidental events, the coincidence timing window was deliberately widened to about \qty{1}{\micro s}, so that multiple accidental-coincidence peaks could be clearly observed.

First, the prompt peak (True + Accidental) was defined by a \( 3\sigma \) range around the peak, as shown in the figure.
Next, five accidental-coincidence peaks on each side (ten in total) were defined as background-selection windows.
Since the prompt peak is expected to contain accidental events similar to those in the side peaks, ten spectra were summed and averaged (divided by 10), and the result was used to estimate the contribution of accidental events included in the prompt peak, as indicated by the red line in Fig.~\ref{fig3}.
By summing and averaging multiple spectra, the statistical uncertainty associated with the background subtraction was reduced.
By subtracting this background component from the prompt spectrum, true coincidence events were successfully extracted, providing a clean dataset for various spectral analyses, including energy-correlation and separation-energy spectra.
\subsection{Energy reconstruction and \( \alpha \)-separation spectrum}
Detailed ion-optics analyses of the GR and LAS spectrometers were first performed, from which the actual focal-plane positions were analytically determined.
Based on these results, the horizontal position at the focal plane was directly related to the particle momentum through the magnetic rigidity of the spectrometer, and the corresponding kinetic energy was obtained from the reconstructed momentum.
Using this procedure, the energies of the scattered proton in GR and the knocked-out \( \alpha \) particle in LAS were determined on an event-by-event basis.

The energy conservation in the \( (p,p\alpha) \) reaction is expressed as
\begin{equation}
   E_{p} + E_{\alpha} + E_{\mathrm{r}}
   =
   E_{\mathrm{beam}} + m_{\mathrm{t}} ,
\end{equation}
where \( E_{p} \) and \( E_{\alpha} \) denote the total energies of the scattered proton and the knocked-out \( \alpha \) particle, and \( E_{\mathrm{r}} \) is the total energy of the residual nucleus.
The target mass \( m_\mathrm{t} \) is given.
Rewriting this equation in terms of kinetic energies and masses gives
\begin{equation}
   T_{\mathrm{beam}} - (T_{p} + T_{\alpha} + T_{\mathrm{r}})
   =
   (m_{\alpha} + m_{\mathrm{r}} - m_{\mathrm{t}})
   \equiv
   S_{\alpha} .
\end{equation}
Here, \( S_{\alpha} \) represents the \( \alpha \)-separation energy, i.e., the energy required to remove an \( \alpha \) particle from the target nucleus.
Since the kinetic energy of the residual nucleus is negligibly small and the beam energy \( T_{\mathrm{beam}} \) is fixed at \qty{392}{MeV}, the following approximation holds for a constant \( S_{\alpha} \):
\begin{equation}
   T_{p} + T_{\alpha} \approx \mathrm{const.}
\end{equation}

\begin{figure}[htbp]
   \centering
   \begin{subfigure}[b]{0.49\linewidth}
      \centering
      \includegraphics[width=\linewidth]{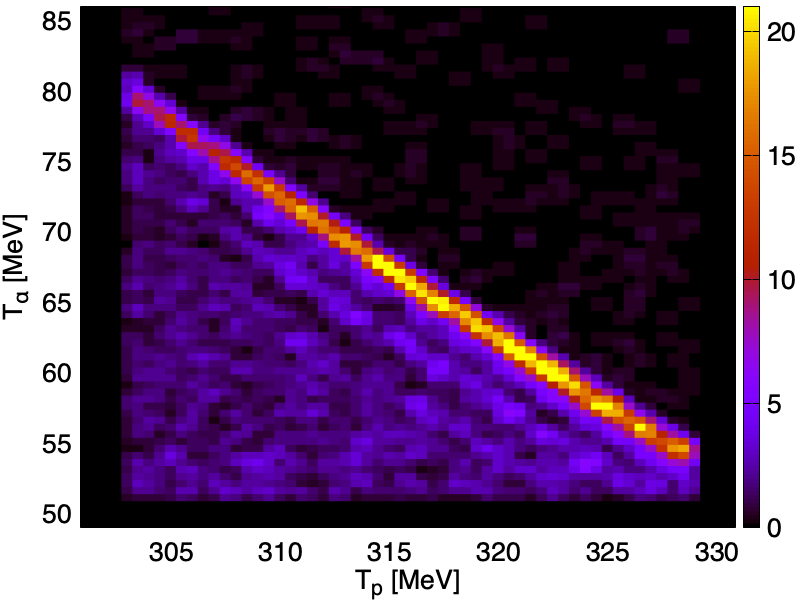}
      \caption{Prompt-peak selection window}
      \label{fig4a}
   \end{subfigure}
   \hfill
   \begin{subfigure}[b]{0.49\linewidth}
      \centering
      \includegraphics[width=\linewidth]{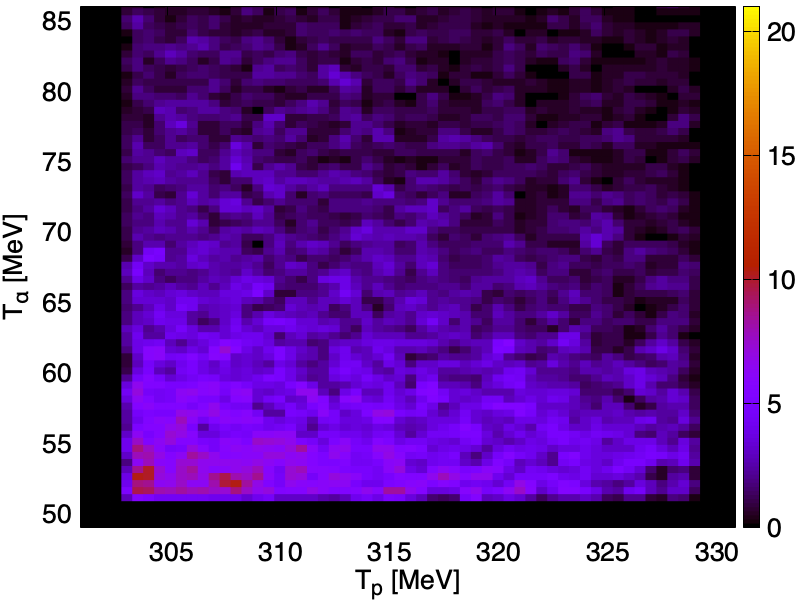}
      \caption{Background selection window}
      \label{fig4b}
   \end{subfigure}
   \caption{
   Energy correlation between the proton and the \( \alpha \) particle in each coincidence window.
   Panel (a) shows the prompt-peak selection, where the events are distributed along a line corresponding to constant \( T_{p} + T_{\alpha} \), demonstrating energy conservation in the \( (p,p\alpha) \) reaction.
   Panel (b) shows the background selection, where the background originating from accidental coincidences exhibits no structure.
   }
   \label{fig4}
\end{figure}

Figure~\ref{fig4}(a) shows the energy correlation between the scattered proton and the knocked-out \( \alpha \) particle, displayed as a density plot with the proton energy on the horizontal axis and the \( \alpha \)-particle energy on the vertical axis.
Only events within the prompt-peak selection window defined in the coincidence timing spectrum are plotted in this figure.
The true coincidence events are distributed along a straight line corresponding to \( T_{p} + T_{\alpha} = \mathrm{const.} \), reflecting energy conservation in the \( (p,p\alpha) \) reaction and confirming that an \( \alpha \) cluster was knocked out from the target nucleus.
In contrast, Fig.~\ref{fig4}(b) shows the events selected by the background selection window, where the background originating from accidental coincidences exhibits no structure.

\begin{figure}[htbp]
   \centering
   \includegraphics[width=0.55\linewidth]{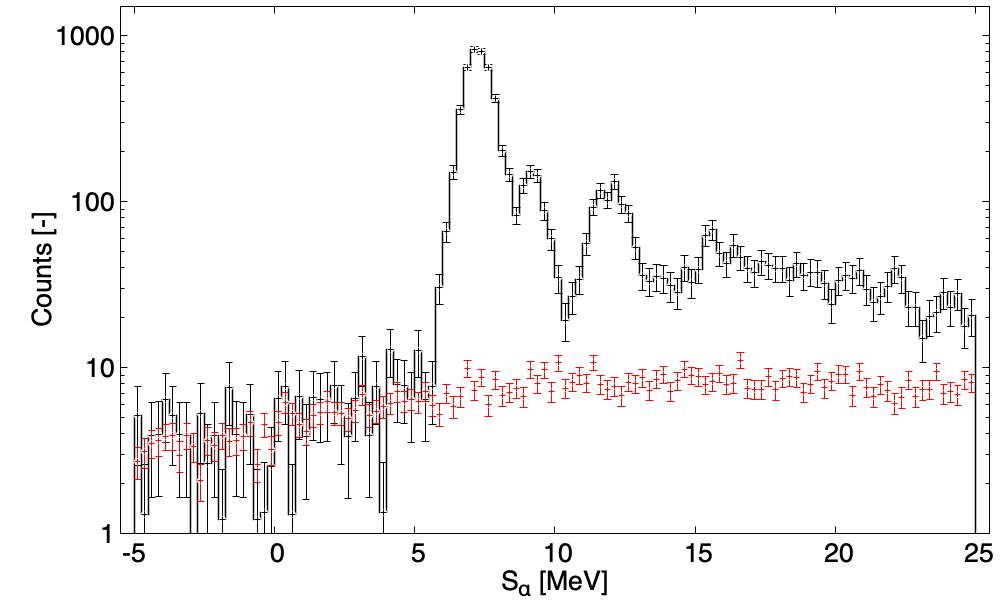}
   \caption{
   \( S_{\alpha} \) energy spectra constructed under two different coincidence timing conditions.
   The black line corresponds to the prompt-peak selection, and the red line shows the background estimated from accidental coincidences.
   }
   \label{fig5}
\end{figure}

Based on the relation between the kinetic energies and \( S_{\alpha} \) given in Eq.~(2), the \( S_{\alpha} \) energy spectrum was derived.
First, the prompt-peak window was selected in the coincidence timing spectrum, and the \( S_{\alpha} \) spectrum constructed under this condition is shown by the black line in Fig.~\ref{fig5}.
In contrast, when the background window was selected and the spectrum was constructed under the same conditions, the result scaled by a factor of 1/10 (corresponding to the average contribution estimated from ten accidental peaks) is shown by the red line.
This component represents a flat background arising from accidental coincidences.
In the subsequent analyses of the \( S_{\alpha} \) spectrum, this red background component was subtracted to isolate the contribution from true coincidence events.
\subsection{Target composition analysis}
Because calcium is highly prone to oxidation, a detailed analysis of the target composition was required.
To quantitatively evaluate the degree of oxidation and possible carbon contamination, a high-precision measurement of proton elastic scattering at \qty{65}{MeV} was performed at the same facility.
This measurement was conducted independently of the \( (p,p\alpha) \) experiment, immediately after it, using one day of beam time.
The Grand Raiden (GR) spectrometer was placed at a laboratory angle of \qty{28}{\degree}.
The measurements were carried out using both the \isotope[\text{nat}]{Ca} target and a Mylar reference target under identical magnetic-field settings.
The \isotope[\text{nat}]{Ca} target used in the elastic-scattering measurement was identical to that employed in the \( (p,p\alpha) \) experiment and had a thickness of \qty{11.8}{mg/cm^2}.
The target thickness was determined from its weight and irradiated area, and the uncertainty of \qty{5}{\%} originates primarily from the precision of the area measurement.
Mylar, whose chemical composition is well known \( (\mathrm{C_{10}H_{8}O_{4}})_n \), was used as a reference.
By weighing a large sheet (\qty{10}{cm} \( \times \) \qty{10}{cm}), its thickness and the elemental composition were determined with high accuracy.
The magnetic field was optimized so that protons elastically scattered from \isotope[16]{O} were centered at the focal plane.
Under these conditions, the GR energy acceptance also covered protons elastically scattered from \isotope[12]{C} and \isotope[40]{Ca}.
By comparing the elastic-scattering yields of \isotope{C} and \isotope{O} observed with the \isotope[\text{nat}]{Ca} target with those measured using the Mylar reference target, the relative abundances of carbon and oxygen in the target were quantitatively evaluated.
Table~\ref{tab:composition} lists the contamination levels of \isotope[12]{C} and \isotope[16]{O} contained in the \isotope[\text{nat}]{Ca} target.
Furthermore, by subtracting these major contaminants from the total target thickness and taking into account the natural isotopic abundance of \isotope[40]{Ca}, the content of \isotope[40]{Ca} in the target was also determined.
\begin{table}[htbp]
   \centering
   \caption{
   Composition of the \isotope[\text{nat}]{Ca} target used in the \( (p,p\alpha) \) experiment, including the contaminant fractions of \isotope[12]{C} and \isotope[16]{O}, the corrected content of \isotope[40]{Ca}, and the corresponding \( \alpha \)-separation energies (\( S_{\alpha} \)) of each nucleus.
   }
   \label{tab:composition}
   \begin{tabular}{lccc}
      \hline\hline
      Nucleus & \( S_{\alpha} \) (\unit{MeV}) & Amount in \isotope[\text{nat}]{Ca} target \\
      \hline
      \isotope[12]{C}  & 7.367 & \qty{0.0371(1)}{mg/cm^2} \\
      \isotope[16]{O}  & 7.162 & \qty{0.326(1)}{mg/cm^2}  \\
      \isotope[40]{Ca} & 7.040 & \qty{11.1(5)}{mg/cm^2}   \\
      \hline\hline
   \end{tabular}
\end{table}

Nuclei with similar \( S_{\alpha} \) values contained in the target can produce overlapping peaks in the \( S_{\alpha} \) spectrum.
Because light nuclei generally exhibit larger  \( (p,p\alpha) \) cross sections than medium-mass nuclei, even a small admixture of \isotope[12]{C} or \isotope[16]{O} can give rise to a noticeable background component.
To accurately evaluate and subtract these background contributions, \( (p,p\alpha) \) measurements were performed using \isotope[\text{nat}]{C} and Mylar targets under the same experimental conditions as for the \isotope[\text{nat}]{Ca} target.
Since the fractions of \isotope[12]{C} and \isotope[16]{O} impurities in the \isotope[\text{nat}]{Ca} target were known, the \( S_{\alpha} \) spectra originating from the  \( (p,p\alpha) \) reactions on these nuclei could be predicted on an absolute scale.
The reference spectra obtained from these measurements were used in Sec.~4 to quantitatively subtract the contamination components and extract the pure \isotope[40]{Ca} contribution.
Specifically, the relative contributions from \isotope[12]{C} and \isotope[16]{O} were evaluated using proton elastic-scattering measurements at 65~MeV performed under identical experimental conditions for the Ca targets and reference Mylar and \isotope[\text{nat}]{C} targets.
Because the elastic-scattering yield ratios between the Ca and reference targets depend only on the relative numbers of O and C atoms, these ratios provide a reaction-model-independent determination of the impurity contents in the Ca targets.
The experimentally determined O and C atom ratios were then directly applied to the 392-MeV \( (p,p\alpha) \) data to predict the corresponding \( S_{\alpha} \) spectra on an absolute scale.
Using these predicted spectra, the O and C contributions were quantitatively subtracted from the measured spectra, enabling the extraction of the pure \isotope[40]{Ca} component.
This procedure avoids assumptions on absolute contamination levels or reaction cross sections and relies solely on experimentally measured yield ratios obtained under identical kinematic and instrumental conditions.
The details of this procedure are introduced in Ref.~\cite{Tanaka2025}.
\subsection{Angular acceptance calibration}
To determine the absolute cross section, it is essential to define the angular acceptance of both the GR and LAS spectrometers.
In the present experiment, no mechanical restrictions such as collimators were applied, and data were taken over the full angular range that each spectrometer can cover.
Based on a detailed analysis of separately obtained angular-calibration data, a software acceptance was defined with careful consideration to avoid any loss of events due to the geometry of the beam duct or the focal-plane detectors.
\section{Results and Discussion}
As described in Sec.~3, coincidence events corresponding to the \isotope[40]{Ca}\( (p,p\alpha) \) reaction were extracted.
Accidental coincidences were identified as a smooth background and removed.
Contaminant components originating from \isotope[12]{C} and \isotope[16]{O} were quantitatively corrected based on reference measurements using \isotope[\text{nat}]{C} and Mylar targets.
Through these procedures, the \( \alpha \)-separation energy (\( S_{\alpha} \)) spectrum of pure \isotope[40]{Ca} was obtained.
The main experimental results are presented below.
\subsection{Alpha-separation energy spectra}
\begin{figure}[htbp]
   \centering
   \begin{subfigure}[b]{0.49\linewidth}
      \centering
      \includegraphics[width=\linewidth]{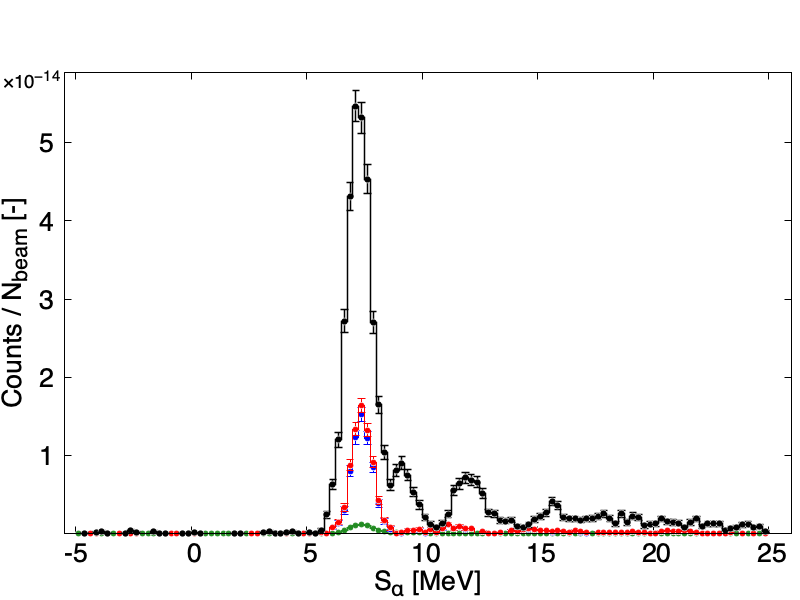}
      \caption{After subtraction of accidental background.}
      \label{fig6a}
   \end{subfigure}
   \hfill
   \begin{subfigure}[b]{0.49\linewidth}
      \centering
      \hspace{-5mm}
      \includegraphics[width=\linewidth]{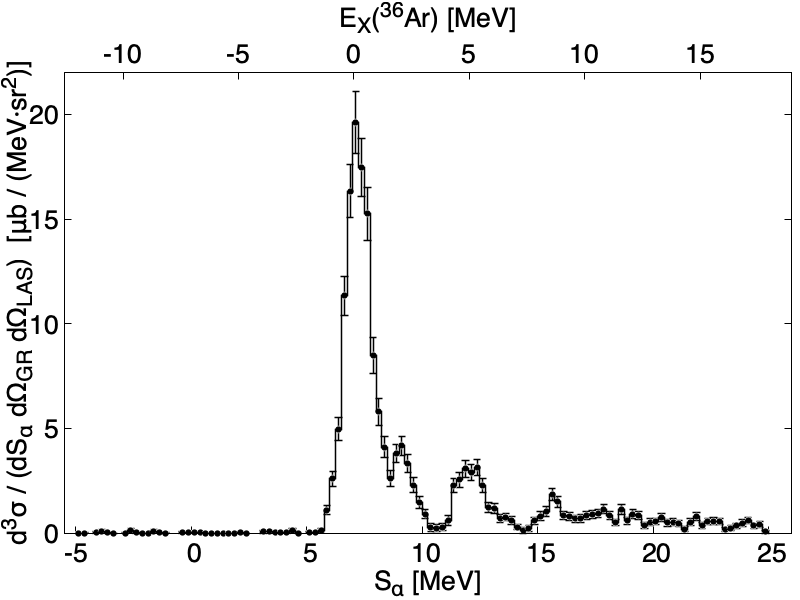}
      \caption{After subtraction of contamination.}
      \label{fig6b}
   \end{subfigure}
   \caption{
   \( \alpha \)-separation energy (\( S_{\alpha} \)) spectra for the \isotope[40]{Ca} target.
   (a) Spectrum after removing accidental background, overlaid with the reference data from \isotope[\text{nat}]{C} and Mylar targets.
   The green curve represents the contribution from carbon, the blue curve corresponds to oxygen contamination, and the red curve shows their sum.
   (b) Extracted \isotope[40]{Ca} spectrum after subtracting oxygen and carbon contributions.
   }
   \label{fig6}
\end{figure}

Figure~\ref{fig6} shows the \( \alpha \)-separation energy (\( S_{\alpha} \)) spectra obtained from the \isotope[40]{Ca} target.
Panel (a) presents the spectrum after subtraction of accidental coincidences, where contributions from contaminant nuclei are visible.
The vertical axis represents the yield normalized to the number of incident protons.
The green and blue histograms correspond to \isotope[12]{C} and \isotope[16]{O}, respectively, and indicate the predicted contributions on an absolute scale, derived from measurements with \isotope[\mathrm{nat}]{C} and Mylar targets under the same \( (p,p\alpha) \) experimental conditions.
As evaluated in Sec.~3, the contamination levels of \isotope[12]{C} and \isotope[16]{O} were small; however, their contributions are non-negligible compared with the \isotope[40]{Ca} component, because the \( (p,p\alpha) \) reaction cross sections for light nuclei are relatively large.
Panel (b) displays the resulting \isotope[40]{Ca} spectrum obtained after subtracting the \isotope[12]{C} and \isotope[16]{O} components.
The vertical axis represents the triple differential cross section, taking into account the target thickness, angular acceptance, and bin width of the \( S_{\alpha} \) spectrum.
Gaussian fits to the \( S_{\alpha} \) peaks yielded an experimental resolution of
\( \sigma = \qty{0.46 \pm 0.01}{MeV} \), where the quoted uncertainty represents the statistical error obtained from the fit covariance matrix.
This resolution can be consistently explained by two dominant contributions.
The variation in the energy loss of the emitted \( \alpha \) particles, depending on the reaction depth within the target, contributes approximately \qty{0.36}{MeV}.
In addition, the beam energy resolution of \qty{0.25}{MeV} contributes to the observed width.
The quadrature sum of these effects quantitatively reproduces the measured resolution.

The \( \alpha \)-separation energy \( S_{\alpha} \) obtained from the \isotope[40]{Ca}\( (p,p\alpha) \) reaction is directly related to the excitation energy \( E_x \) of the residual nucleus \isotope[36]{Ar} through the following relation:
\begin{equation}
   E_x \big( \isotope[36]{Ar} \big)
   =
   S_{\alpha} - 7.040 \text{ MeV }.
\end{equation}
Here, \qty{7.040}{MeV} represents the \( \alpha \)-separation energy required to remove an \( \alpha \) cluster from the ground state of \isotope[40]{Ca}, leaving the residual nucleus \isotope[36]{Ar} in its ground state.
In the framework of quasi-free scattering (QFS), it is assumed that the target nucleus is initially in its ground state before the reaction.
Therefore, this separation energy serves as a reference for defining the excitation energy of the residual nucleus.
The most prominent peak in the spectrum corresponds to the transition to the ground state of \isotope[36]{Ar}, while the adjacent peak on the higher-energy side corresponds to the first excited state at \qty{1.97}{MeV}, which is resolved with a statistical significance of about \( 2\sigma \).
At higher excitation energies, around \qty{5}{MeV}, several closely spaced states---including \( 2^+ \) (\qty{4.951}{MeV}), \( 2^- \) (\qty{4.974}{MeV}), and \( 5^- \) (\qty{5.171}{MeV})---are expected to overlap, leading to the formation of a broad composite structure rather than distinct peaks.
Because a natural \isotope{Ca} target was used, minor \isotope{Ca} isotopes are present; however, their contributions do not affect the present analysis.
In particular, the \( S_{\alpha} \) value of \isotope[44]{Ca} (\qty{8.854}{MeV}) differs from that of \isotope[40]{Ca} (\qty{7.040}{MeV}) by \qty{1.815}{MeV}, which is significantly larger than the \( \pm 2\sigma \) selection window (\qty{\pm 0.92}{MeV}) applied to the \isotope[40]{Ca} ground-state peak.
As a result, the tail of the \isotope[44]{Ca} contribution does not enter the \isotope[40]{Ca} gate, and its effect can be neglected in view of its small isotopic abundance (\qty{2.086}{\%}).
\subsection{Energy distribution of the emitted proton}
\begin{figure}[htbp]
\centering
\begin{subfigure}[b]{0.49\linewidth}
   \centering
   \includegraphics[width=\linewidth]{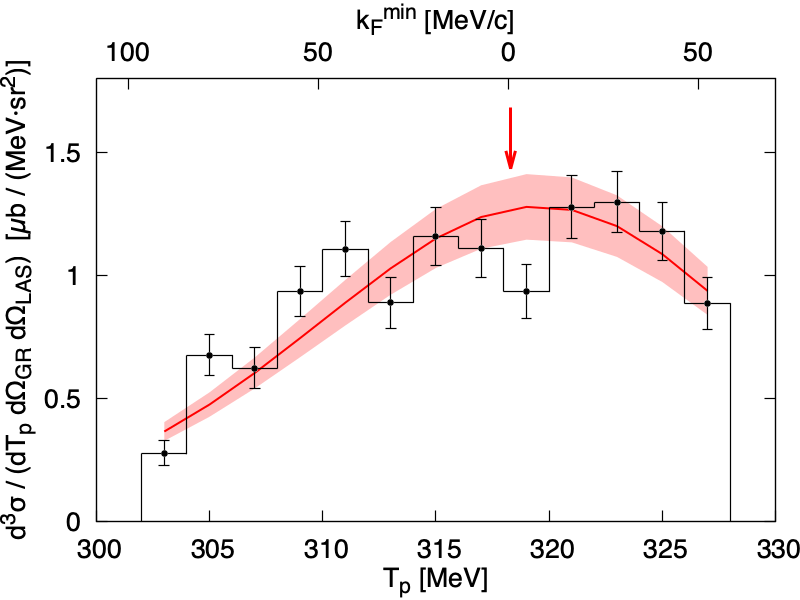}
   \caption{
   Proton energy distribution of present data at \qty{392}{MeV} (RCNP).
   }
   \label{fig7a}
\end{subfigure}
\hfill
\begin{subfigure}[b]{0.49\linewidth}
   \centering
   \hspace{-5mm} 
   \includegraphics[width=\linewidth]{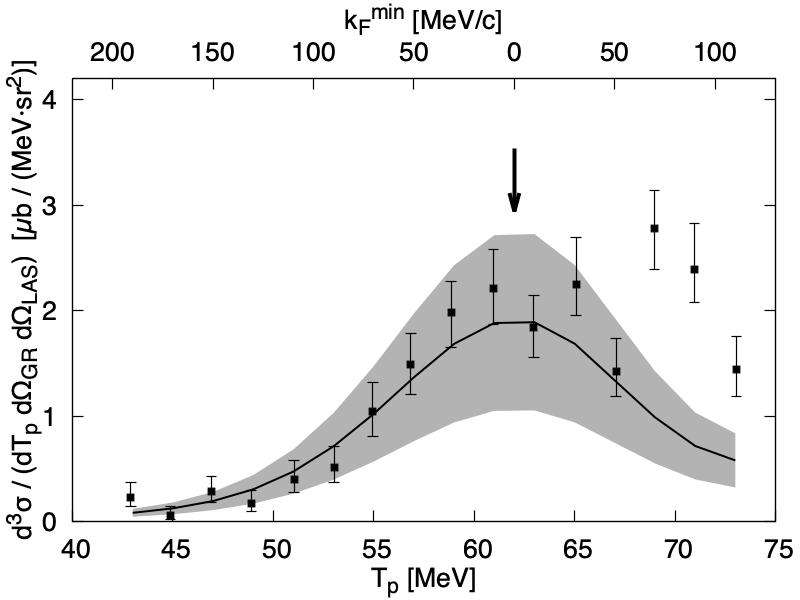}
   \caption{Proton energy distribution of previous data at \qty{101.5}{MeV} (Maryland Cyclotron)~\cite{Carey1981,Carey1984}.}
   \label{fig7b}
\end{subfigure}
\caption{
   Proton energy distributions for the \( \isotope[40]{Ca}(p,p\alpha) \) reaction.
   The vertical axis represents TDX \( d^3\sigma/(dT_{p}\, d\Omega_{p}\, d\Omega_{\alpha}) \).
   The curves are the DWIA results scaled to the experimental data to minimize the \( \chi^2 \), and the hatched areas represent the uncertainties of the fit.
   Error bars indicate statistical uncertainties only.
   }
\label{fig7}
\end{figure}

To investigate the energy distribution characteristic of each state, events in the \( \alpha \)-separation energy region corresponding to the ground-state transition \( \isotope[40]{Ca}(0^+) \to \isotope[36]{Ar}(0^+) \) were selected within a range of \( \pm 2\sigma \).
The energy distribution of the emitted proton, measured with the GR spectrometer, is shown in Fig.~\ref{fig7}(a).
All background contributions discussed earlier have been removed.

The vertical axis represents the triple differential cross section (TDX), defined as
\begin{equation}
   \frac{d^3\sigma}{dT_{p}\, d\Omega_{p}\, d\Omega_{\alpha}}
   =
   \frac{Y}{N_{\mathrm{beam}} N_{\mathrm{target}}\, \Delta T_{p}\, \Delta \Omega_{p}\, \Delta \Omega_{\alpha}} ,
\end{equation}
where \( Y \) is the yield of detected \( p\text{--}\alpha \) coincidence events, \( N_{\mathrm{beam}} \) is the total number of incident protons, and \( N_{\mathrm{target}} \) is the areal density of target nuclei.
\( T_{p} \) denotes the kinetic energy of the scattered proton, while \( \Omega_{p} \) and \( \Omega_{\alpha} \) represent the solid angles subtended by the proton and \( \alpha \)-particle detectors, respectively.
All relevant detection efficiencies and normalization factors were taken into account.
In a knockout reaction, the reaction kinematics can be uniquely determined by specifying \( T_{p} \), \( \Omega_{p} \), and \( \Omega_{\alpha} \).
The corresponding reaction rate is expressed as the TDX, which serves as a physical quantity directly comparable with theoretical calculations at fixed kinematics.

For the transition from the ground state (\( 0^{+} \)) of the target nucleus to the ground state (\( 0^{+} \)) of the residual nucleus, no orbital angular momentum transfer is involved.
Under this condition, the reaction is expected to reach its maximum cross section when the recoil momentum of the residual nucleus is minimized, i.e., under the recoil-less condition.
In Fig.~\ref{fig7}(a), the upper horizontal axis shows the minimum recoil momentum \( k_{F}^{\rm min} \) of the residual nucleus that can be accessed within the angular acceptance, as a function of the proton energy.
From momentum conservation in the initial state and the quasi-free condition, the recoil momentum of the residual nucleus is related to the internal momentum of the knocked-out \( \alpha \) particle by \( \vec{k}_{F} = -\vec{k}_{\alpha} \).
According to the kinematical calculation, the recoil-less condition in Fig.~\ref{fig7}(a) corresponds to the arrow at \( T_{p} = \qty{320}{MeV} \).
The measured triple-differential cross-section (TDX) spectrum indeed exhibits a clear maximum around this energy, consistent with the expectation for a recoil-less transition.
\subsection{Comparison with previous data and reaction calculations}
To examine the consistency of the present measurement, the results were compared with the existing \( \isotope[40]{Ca}(p,p\alpha) \) data taken at \qty{101.5}{MeV}~\cite{Carey1981,Carey1984}.
This dataset represents one of the most thoroughly analyzed and reliable measurements among previous studies.
Figure~\ref{fig7}(b) shows the energy distribution of the scattered protons.
This distribution is compared with the present \qty{392}{MeV} data.
Because the reaction kinematics, the elementary \( p\text{--}\alpha \) scattering cross section, and the strength of refraction and absorption effects differ between the two energy regimes, a direct comparison of the absolute TDX values is not straightforward.
Therefore, detailed reaction calculations are required to perform a meaningful comparison and to assess the consistency of the underlying reaction mechanism.
\subsection{DWIA calculation}
\label{subsec:DWIA}
To interpret the measured cross sections, DWIA (Distorted Wave Impulse Approximation) calculations were performed using the computer code \textsc{Pikoe}~\cite{Ogata2024}.
The \( \alpha \)-cluster wave function in \( \isotope[40]{Ca} \) was obtained as the bound-state wave function of the \( \alpha + \isotope[36]{Ar} \) system.
A Woods--Saxon form was assumed for the binding potential, and its depth was adjusted to reproduce the empirical \( \alpha \)-separation energy.
The radius and diffuseness parameters were set to \( R_0 = 1.25 \times 36^{1/3}\,\unit{fm} \) and \( a_0 = 0.76\,\unit{fm} \), respectively.
These values were determined in the \( \alpha \)-transfer reaction analysis by Fukui \textit{et al.}~\cite{Fukui2016}, in which the \( \alpha \)-cluster wave function was adjusted to reproduce, as closely as possible, the shape of the microscopic cluster model.
The obtained \( \alpha \)-cluster wave function was normalized to unity.
This part corresponds to the structural preparation, which is used in the subsequent reaction-theoretical analysis.

In the DWIA framework for knockout reactions, the differential cross section of the \( p\text{--}\alpha \) elementary process is used as an input.
For the entrance and exit channels, the EDAD1 global Dirac phenomenological optical potential~\cite{Hama1990,Cooper1993,Cooper2009} was adopted for the \( p + \isotope[40]{Ca} \) and \( p + \isotope[36]{Ar} \) systems, while a global \( \alpha \)--nucleus optical potential from Ref.~\cite{Avrigeanu1994} was employed for the \( \alpha + \isotope[36]{Ar} \) interaction.
The \( p\text{--}\alpha \) elementary cross section is also obtained using the optical potential by the Dirac-phenomenology~\cite{Cooper2009} consistently with the optical potentials for the distorted waves.
We also confirmed that a direct use of the elastic scattering data at \qty{85}{MeV}~\cite{Votta1974} and \qty{350}{MeV}~\cite{Moss1980} for \qty{101.5}{MeV} and \qty{392}{MeV} cases, respectively, makes little difference in the extracted \( S_\mathrm{FAC} \).
Using these optical potentials and \( p\text{--}\alpha \) cross section, the TDXs as a function of proton energy were calculated.

The calculations were performed by sampling the scattering angles within the experimental angular acceptance and taking their weighted average to reproduce the actual acceptance of the spectrometer.
This treatment explicitly accounts for the finite angular acceptance, which leads to a slight broadening of the calculated energy distribution.

As discussed in Eqs.~(3.52--3.57) of Ref.~\cite{Wakasa2017} and also in other DWIA references therein, TDX is essentially proportional to the absolute square of the distorted momentum distribution of the \( \alpha \) amplitude.
Pure momentum distribution, which is the Fourier transform of the \( \alpha \) amplitude in the coordinate space, is relevant to the TDX once the plane-wave impulse approximation (PWIA) is adopted.
Therefore, the reproduction of the data in Fig.~\ref{fig7} shows that the product of the \( \alpha \) amplitude and three distorted waves and thus the distorted momentum distribution are described properly.
To quantitatively evaluate the effects of the distortion, PWIA calculations were also performed and discussed in the following analysis.
\begin{table}[htbp]
   \centering
   \caption{
   Comparison of PWIA and DWIA calculations under recoilless kinematics.
   PWIA \( \mathrm{TDX_{calc}} \) and DWIA \( \mathrm{TDX_{calc}} \) are computed with the \( \alpha \)-cluster wave function normalized to unity.
   The input \( p\text{--}\alpha \) differential cross sections (DX) are taken from experimental data at 85~MeV~\cite{Votta1974} (for the \qty{101.5}{MeV} case) and \qty{350}{MeV}~\cite{Moss1980} (for the \qty{392}{MeV} case).
   The quoted values correspond to the recoilless condition at a CM scattering angle of \qty{85}{\degree} (for \qty{101.5}{MeV}) and \qty{60}{\degree} (for \qty{392}{MeV}).
   Calculations were performed using \textsc{Pikoe}~\cite{Ogata2024}.
   The extracted experimental spectroscopic factors \( S_{\mathrm{FAC}}^{\mathrm{WS}} \) from \( \isotope[40]{Ca}(p,p\alpha) \) are also listed.
   Uncertainties combine statistical and fitting (stat.\&fit.) and systematic (syst.) contributions.
   }
   \label{tab:dwia}
   \begin{tabular}{lcc}
      \hline\hline
       & \textbf{\qty{101.5}{MeV}} & \textbf{\qty{392}{MeV}} \\
      \hline
      \( p\text{--}\alpha \) DX$_{\mathrm{exp}}$ (mb/sr)                  & 2.31                     & 0.0931                 \\
      PWIA \( \mathrm{TDX_{calc}} \) ($\mu$b/(MeV$\cdot$sr$^{2}$))       & 257.9                    & 47.1                   \\
      DWIA \( \mathrm{TDX_{calc}} \) ($\mu$b/(MeV$\cdot$sr$^{2}$))       & 4.5                      & 2.6                    \\
      $R_{\mathrm{PWIA/DWIA}}$                                           & 57.3                     & 18.1                   \\
      TDX$_{\mathrm{exp}}$ ($\mu$b/(MeV$\cdot$sr$^{2}$))                 & $\sim$2.3                & $\sim$1.3              \\
      $S_{\mathrm{FAC}}^{\mathrm{WS}}$ (--)                              & 0.52                     & 0.51                   \\
      $\Delta_{\mathrm{stat.\&fit.}}S_{\mathrm{FAC}}^{\mathrm{WS}}$ (--) & $\pm$0.23                & $\pm$0.05              \\
      $\Delta_{\mathrm{syst.}}S_{\mathrm{FAC}}^{\mathrm{WS}}$ (--)       & $\pm$0.05                & $\pm$0.06              \\
      \hline\hline
   \end{tabular}
\end{table}

To quantify the impact of refraction and absorption, the PWIA-to-DWIA ratio is defined as
\begin{equation}
   R_{\mathrm{PWIA/DWIA}} \equiv \mathrm{PWIA}/\mathrm{DWIA}.
\end{equation}
For the present systems, the obtained values are \( R_{\mathrm{PWIA/DWIA}} = 57.3 \) at \qty{101.5}{MeV} and \( R_{\mathrm{PWIA/DWIA}} = 18.1 \) at \qty{392}{MeV}.
This result indicates that the PWIA significantly overestimates the absolute cross sections by one to two orders of magnitude compared with the DWIA.
Such large PWIA-to-DWIA ratios are not specific to the \( (p,p\alpha) \) reaction; for example, a typical value of \( R_{\mathrm{PWIA/DWIA}} \approx 25 \) has been reported for the \( \isotope[208]{Pb}(p,2p) \) reaction at \qty{200}{MeV}~\cite{Cowley2020}.
The pronounced energy dependence of \( R_{\mathrm{PWIA/DWIA}} \) demonstrates that the absorption effect becomes weaker at higher incident energies.
Specifically, when the energy increases from \qty{101.5}{MeV} to \qty{392}{MeV}, the absorption is reduced by approximately \qty{68}{\%} (\( 1 - 18.1/57.3 \approx 0.68 \)).
This trend is consistent with the general understanding that the distorting potentials become less absorptive as the beam energy increases.
Therefore, when comparing the absolute TDX values across different incident energies, it is necessary to take into account this energy-dependent absorption effect in addition to the dependence on the CM scattering angle.
\subsection{Experimental spectroscopic factors of \( \alpha \)-clusters in the Woods--Saxon model}
\label{subsec:N-fac}
The experimental spectroscopic factor of \( \alpha \)-clusters based on the Woods--Saxon model, denoted as \( S_{\mathrm{FAC}}^{\mathrm{WS}} \), was extracted by comparing the TDX calculated using DWIA, \( \mathrm{TDX_{calc}} \), with the experimentally measured value, \( \mathrm{TDX_{exp}} \).
It is defined as
\begin{equation}
   S_{\mathrm{FAC}}^{\mathrm{WS}}
   =
   \frac{\mathrm{TDX_{exp}}}{\mathrm{TDX_{calc}}}.
\end{equation}
Because the shapes of the proton-energy (\( T_{p} \)) distributions for \( \mathrm{TDX_{exp}} \) and \( \mathrm{TDX_{calc}} \) are not identical, \( S_{\mathrm{FAC}}^{\mathrm{WS}} \) was determined by minimizing the chi-squared difference between the two spectra, as shown in Fig.~\ref{fig7}.
The solid line in the figure represents the calculated \( \mathrm{TDX_{calc}} \) corresponding to the minimum point-wise \( \chi^{2} \), while the shaded band indicates the uncertainty range associated with \( \Delta\chi^{2} = 1 \), corresponding to the \( 1\sigma \) confidence interval.
The extracted values of the experimental spectroscopic factor are summarized in Table~\ref{tab:dwia}.
Remarkably, despite the large difference in incident energies and experimental conditions, the extracted experimental spectroscopic factors at \qty{101.5}{MeV} and \qty{392}{MeV} are consistent within uncertainties.
This agreement demonstrates that the present high-energy measurement reproduces the same underlying reaction mechanism as the well-established \qty{101.5}{MeV} experiment, thereby confirming the reliability of \( (p,p\alpha) \) studies in the several-hundred-MeV region.
We therefore conclude that the \( (p,p\alpha) \) measurement at \qty{392}{MeV} has been successfully validated.
This achievement paves the way for systematic studies under a broader range of kinematic conditions than those accessible around \qty{100}{MeV}, allowing detailed exploration of the intrinsic momentum distribution of \( \alpha \) clusters and providing new insight into their density distributions near the nuclear surface.

It should be noted, however, that the phenomenological \( \alpha \)-cluster wave function obtained using a Woods--Saxon potential tends to overestimate the \( \alpha \) amplitude in the nuclear interior (see, e.g., Fig.~11 of Ref.~\cite{Carey1984} and Fig.~1 of Ref.~\cite{Yoshida2019}).
This overestimation arises because the Woods--Saxon model does not explicitly incorporate the Pauli exclusion principle (antisymmetrization) between the nucleons in the \( \alpha \) cluster and those in the core.
Since the \( (p,p\alpha) \) reaction is strongly peripheral owing to absorption effects, DWIA analyses based on such phenomenological wave functions---including the present study---may systematically overestimate the absolute value of the experimental spectroscopic factor \( S_{\mathrm{FAC}}^{\mathrm{WS}} \).
To address the effect quantitatively, further theoretical investigation is necessary, on both surface sensitivity of the reaction and the suppression on the \( \alpha \) amplitude in the internal region due to the antisymmetrization.
Consistency among theoretical predictions on the strength of the suppression should be also settled.
See Ref.~\cite{Sargsyan2025} for \isotope[20]{Ne} case for example.

Although microscopic cluster models such as OCM and GCM have qualitatively discussed the \( \alpha \)-cluster component in \isotope[40]{Ca}(e.g., Refs.~\cite{Sakuda1998}), fully microscopic approaches—for example AMD or large-scale shell-model calculations—have so far been limited in providing quantitative estimates of the absolute \( \alpha \)-spectroscopic factor.
Further theoretical developments along these lines would be valuable for a more direct connection between reaction observables and nuclear-structure models.
\section{Summary and Conclusions}
In this study, the \( \isotope[40]{Ca}(p,p\alpha)\isotope[36]{Ar} \) reaction was measured using a \qty{392}{MeV} proton beam, establishing a robust experimental framework for quasi-free scattering at high energies.
Whereas previous experiments performed near \qty{100}{MeV}~\cite{Carey1984} provided important insights into \( \alpha \)-cluster formation, the present work extends these achievements by realizing kinematic conditions that more ideally satisfy the quasi-free scattering requirement in the several-hundred-MeV region.

The \( \alpha \)-separation energy (\( S_{\alpha} \)) spectrum has been reconstructed, which directly corresponds to the excitation spectrum of \isotope[36]{Ar} residue, and the transitions to both the ground and excited states were clearly identified.
For the ground-state transition, the measured energy distributions of the outgoing proton were obtained.
The \( \alpha \) spectroscopic factor extracted by the present DWIA analysis was consistent with the former high-precision data at \qty{100}{MeV}~\cite{Carey1984}, across widely different beam energies.
This agreement demonstrates that the underlying reaction mechanism is common to both low- and high-energy regimes, thereby confirming the reliability of \( (p,p\alpha) \) studies in the several-hundred-MeV region.

The achieved energy resolution was sufficient to resolve individual excited states, demonstrating the potential for future state-by-state analyses.
Since the energy and momentum distributions of the \( (p,p\alpha) \) reaction reflect the momentum distribution of the \( \alpha \) clusters and thus its spatial distribution, extending measurements to a broader range of scattering angles and kinematic conditions will be crucial for mapping their distribution near the nuclear surface.
The results extend the knowledge obtained at \( \sim \qty{100}{MeV} \) to higher energies under more favorable kinematic conditions, establishing a firm experimental foundation for systematic studies of \( \alpha \) clustering in both stable and unstable nuclei~\cite{Tanaka2025-2}.

\section*{Acknowledgments}
This work was performed under the research project KK-002 of the Research Center for Nuclear Physics, the University of Osaka.
The authors would like to express their sincere gratitude to the accelerator group at the Research Center for Nuclear Physics (RCNP), Sumitomo Heavy Industries Accelerator Service Ltd. (SAS), and all supporting staff for providing the high-quality proton beam and their invaluable technical assistance during the experiments.
This work was supported by JSPS KAKENHI (Grant Number JP21H04975) and was partially supported by the JSPS A3 Foresight Program ``Nuclear Physics in the 21st Century'', also in part by the National Key R\&D Program of China (Grants No. 2023YFE0101500, No. 2022YFA1605100), the National Natural Science Foundation of China (Grant No. 12275006).
This work was also partly supported by the Institute for Basic Science (IBS) funded by the Ministry of Science and ICT, Korea (Grants No. IBS-R031-D1), and the National Research Foundation of Korea (NRF) (Grant No. RS-2024-00333673).
This work was also partly supported by the UK Science and Technology Facilities Council (STFC) (Grants ST/V001035/1 and ST/Y000285/1).
%J.T. sincerely thanks Nori Aoi and Mitsuhiro Fukuda for their patient support and the accelerator facility upgrade.


%\bibliographystyle{ptephy}
%\bibliography{sample}
%
% once the .bbl file has been generated then place the text in your article.

\vspace{0.2cm}
\noindent
%For references,  note how to include DOI information from examples below. 

%This is added by T. Yoneya (editor-in-chief) on 2020/07/09.

\let\doi\relax

%without this code before the command "\begin{thebibliography}{}" , an error will be %flagged. When the bibliography is provided as separate .bib file, then this code %should be placed above the commands "\bibliographystyle{}" and "\bibliography{}" %inside the main TeX file. 

%\bibitem{Taniguchi2021}
%Y.~Taniguchi, K.~Yoshida, Y.~Chiba, Y.~Kanada-En'yo, M.~Kimura, and K.~Ogata,
%Phys.\ Rev.\ C \textbf{103}, L031305 (2021).

%\bibitem{Chant1977}
%N.~S.~Chant and P.~G.~Roos,
%Phys.\ Rev.\ C \textbf{15}, 57 (1977).

%\bibitem{Chant1983}
%N.~S.~Chant and P.~G.~Roos,
%Phys.\ Rev.\ C \textbf{27}, 1060 (1983).

%\bibitem{Nadasen1981}
%A.~Nadasen, P.~G.~Roos, N.~S.~Chant, A.~A.~Cowley, C.~Samanta, and J.~Wesick,
%Phys.\ Rev.\ C \textbf{23}, 2353 (1981).

%\bibitem{Cowley2008}
%A.~A.~Cowley, S.~V.~F\"ortsch, J.~J.~Lawrie, \textit{et al.},
%Phys.\ Rev.\ C \textbf{77}, 027601 (2008).

%\bibitem{Cowley1994}
%A.~A.~Cowley, G.~F.~Steyn, S.~V.~F\"ortsch, J.~J.~Lawrie, J.~V.~Pilcher, F.~D.~Smit, and D.~M.~Whittal,
%Phys.\ Rev.\ C \textbf{50}, 2449 (1994).

%\bibitem{Jain2008}
%A.~K.~Jain and B.~N.~Joshi,
%Phys.\ Rev.\ C \textbf{77}, 027601 (2008).

%\bibitem{Yoshida2018}
%K.~Yoshida, K.~Ogata, and Y.~Kanada-En’yo,
%Phys.\ Rev.\ C \textbf{98}, 024614 (2018).

%\bibitem{Shim2023}
%S.~I.~Shim, K.~Yoshida, and K.~Ogata,
%J.\ Phys.\ Soc.\ Jpn.\ \textbf{92}, 094201 (2023).

\end{document}